\documentclass[]{revtex4}
\usepackage[utf8]{inputenc}
\usepackage[T1]{fontenc}
\usepackage{xcolor}
\usepackage{graphicx}

\begin{document}
\title{Interevent time distributions of avalanche dynamics}
\author{Pinaki Kumar$^{1}$, Evangelos Korkolis$^{2}$, Roberto Benzi$^{3}$, Dmitry Denisov$^{4}$, Andr\'e Niemeijer$^{2}$, Peter Schall$^{4}$, Federico Toschi$^{1,5}$, Jeannot Trampert$^{2}$}
\affiliation{$^{1}$ Department of Applied Physics, Eindhoven University of Technology, P.O. Box 513, 5600 MB Eindhoven, The Netherlands}
\affiliation{$^{2}$ Department of Earth Sciences, Utrecht University, P.O. Box 80115, 3508 TC Utrecht, The Netherlands}
\affiliation{$^{3}$ Dip. di Fisica and INFN, Universit\`a ``Tor Vergata'', Via della Ricerca Scientifica 1, I-00133 Roma, Italy}
\affiliation{$^{4}$ Institute of Physics, University of Amsterdam, 1098 XH Amsterdam, The Netherlands}
\affiliation{$^{5}$ Department of Mathematics and Computer Science,
  Eindhoven University of Technology, P.O. Box 513, 5600 MB Eindhoven,
  The Netherlands and Istituto per le Applicazioni del Calcolo, Consiglio
  Nazionale delle Ricerche, Via dei Taurini 19, 00185 Rome, Italy}



\begin{abstract}
Physical systems characterized by stick-slip dynamics often display avalanches. Regardless of the diversity of their microscopic structure, these systems are governed by a power-law distribution of avalanche size and duration. Here we focus on the interevent times between avalanches and show that, unlike their distributions of size and duration, the interevent time distributions are able to distinguish different mechanical states of the system, characterized by different volume fractions or confining pressures. We use experiments on granular systems and numerical simulations of emulsions to show that systems having the same probability distribution for avalanche size and duration can have different interevent time distributions. Remarkably, for large packing ratios, these interevent time distributions look similar to those for earthquakes and are indirect evidence of large space-time correlations in the system. Our results therefore indicate that interevent time statistics are more informative to characterize the dynamics of avalanches.
\end{abstract}
\maketitle

\flushbottom

%
%
\thispagestyle{empty}


\section*{Introduction}


Many physical systems subject to small external driving forces exhibit complex burst dynamics in space and in time \cite{Fisher,Sethna,Zapperi,uhl_etal:2015,maass_etal:2015}. Burst events are the signature of energy release in the system and, in many cases, they have successfully been described in terms of avalanche dynamics. Both theoretically and experimentally, the statistical distribution of the size, $S$, and the time duration, $t_E$, of avalanches have been shown to satisfy well-defined scaling laws. In the case of plasticity of soft glasses and fracture dynamics of amorphous solids, the scaling exponents are mostly independent of the details of the microscopic interactions, suggesting some form of universality \cite{BenZion,Goldenfeld,Wyart,uhl_etal:2015,Denisov17}. Theoretically, the scaling exponents have been explained by a number of different mean- and non mean-field theories, which take into account the basic physical properties of the system and its intrinsic ``randomness'' \cite{Fisher,BenZion,Wyart}. This apparent universality might also explain the observed scaling properties seen in earthquake dynamics, e.g. the well-known Gutenberg-Richter law \cite{gb:1954,BenZion,uhl_etal:2015}. 
The question naturally arises whether some statistical properties of avalanches are able to discriminate between different states of these systems. One interesting quantity is the interevent time, $t_i$, (also referred to as recurrence time, return time or waiting time) between two consecutive avalanches \cite{maass_etal:2015}. Most of our information on the interevent time distribution $P(t_i)$ comes from the analysis of earthquake catalogs. There is a general consensus that for earthquakes, $P(t_i)$ can be characterized by two different time scales: at short times $P(t_i) \sim 1/t_i$ follows Omori's law \cite{utsu_etal:1995}, while at very long time scales $P(t_i) \sim exp(-\alpha t_i)$. The exponential behaviour is explained by the (reasonable) assumption, that for long time scales, earthquake events are likely to be independent. Corral \cite{Corral2004}, however, showed that $P(t_i)$, for long $t_i$, is better fitted by a Gamma distribution, leading to a time scale defined by the average interevent time.
Interestingly, this Gamma distribution is observed for many different geographic regions, as well as for the whole earth, once $P(t_i)$ is re-scaled to the regional or global average interevent time \cite{Corral2004}. A detailed investigation of seismicity induced by mining and fluid injection revealed the same Gamma distribution for interevent times \cite{Davidsen,Davidsen2} as well as acoustic emissions of various other systems \cite{Niccolini,Ribeiro}. If the interevent time distribution is independent of any size threshold or region, the only possible shape for $P(t_i)$ is exponential, unless complex space and time correlations are present and the system is close to criticality \cite{Molchan,Corral2006}. Another option is that the Gamma distribution emerges as a superposition of independent probability distributions, one with an exponential tail and one with the Omori short-time behaviour~\cite{Saichev2006,Touati}. In the case of superposition, $P(t_i)$ does not reveal any new physics besides the well documented avalanche scaling laws (Gutenberg-Richter and Omori). Finite detection thresholds \cite{Santucci2016} have also been suggested to explain the emergence of Gamma distributions. It is thus a fundamental question to know whether $P(t_i)$ informs us on long-range correlations in the system or not. In the past, studies have tried to settle this question by analyzing the fit of data to various functional forms for $P(t_i)$. Without the knowledge of the underlying state of the system, best fit arguments are obviously difficult to make. We will instead analyze systems where we control the mechanical state (e.g. rigidity) and investigate whether $P(t_i)$ informs us on physics beyond that of $P(S)$ and $P(t_E)$.  

We report on a systematic study of the distribution of interevent times, $P(t_i)$, across different model systems to show that it provides crucial information on the mechanical state of the material, namely the packing ratio or its confining pressure. We complement experimental measurements on granular systems with numerical simulations on emulsion-like models, and compare the resulting interevent time distributions with those for earthquakes. Our results uniquely show that unlike the statistical properties of avalanche sizes and durations, the probability of interevent times $P(t_i)$ strongly depends on the packing ratio or confining pressure of the material. For relatively low packing ratios, the interevent time distribution is exponential with an Omori behaviour at small time scales. For high packing ratios, $P(t_i)$ is similar to that observed for earthquakes at long time scales \cite{Corral2004}, with an Omori scaling at small time scales. This implies that the functional shape of $P(t_i)$ depends on the mechanical properties of the system and that, in addition to avalanche size and duration, it provides a crucial measure to distinguish avalanche-like relaxation mechanisms. We show that spatio-temporal correlations are responsible for these different regimes, and therefore interevent time distributions provide insight into the nature of correlated deformations in dense suspensions and earthquakes.

\section*{Results}

\begin{figure}
\centering
\includegraphics[width=\linewidth]{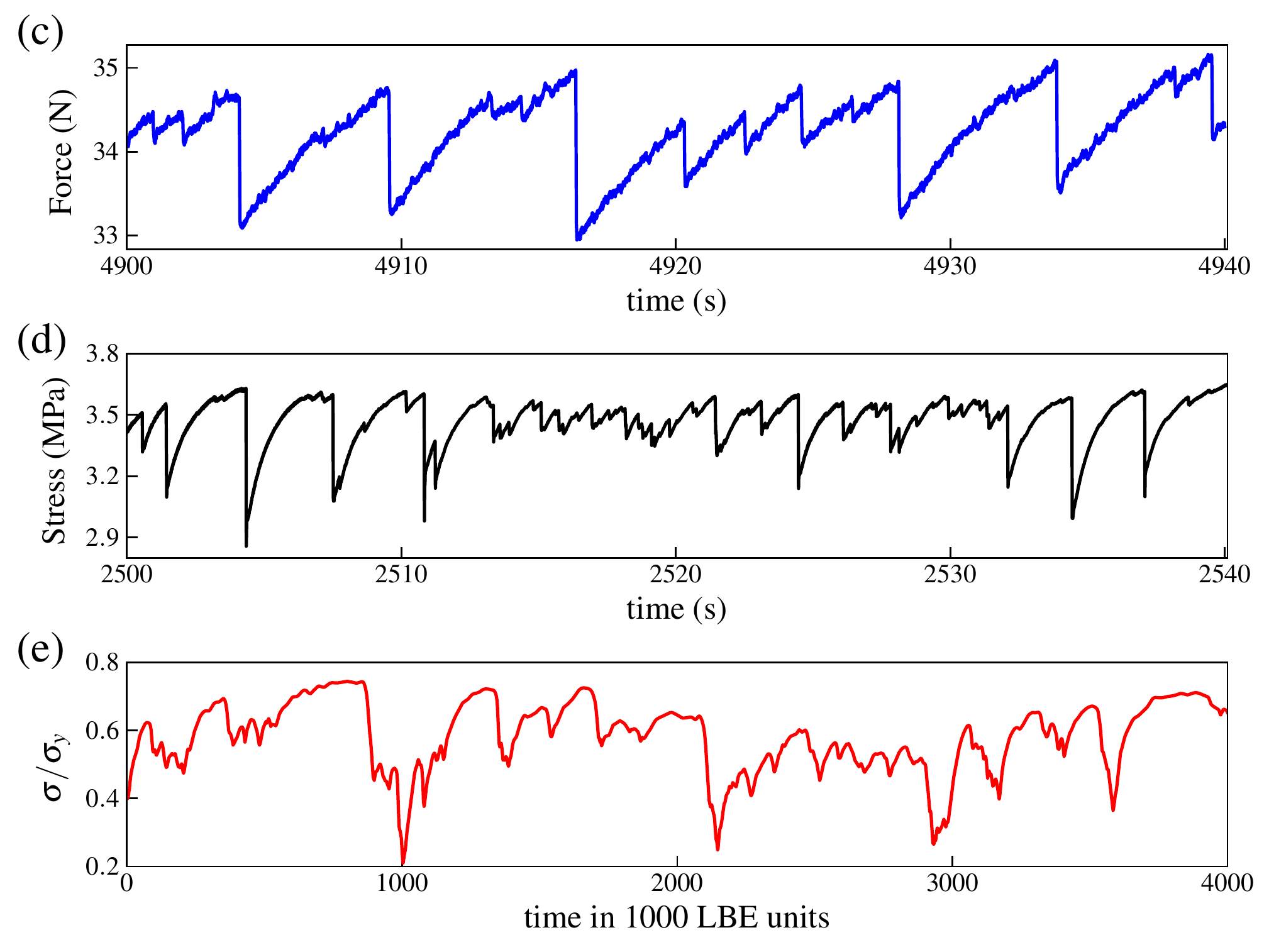}
\caption{Snapshot of the stress/force-time evolution for the different systems. Top: Example from the shear cell, middle: example from the rotary shear experiment and bottom: Results from the LBE simulation for $R=0.09$.}
\label{figure_stress}
\end{figure}

We use granular systems under well-controlled normal forces, apply slow shear strain rates to induce avalanches and monitor them with high temporal resolution (see Methods section).  The recorded force signal from our shear cell exhibits strong intermittency: force increases are followed by sudden force drops that demarcate energy release events (Fig. \ref{figure_stress}~top). We measure the force drops with high temporal resolution to resolve the dynamics of both the large and small avalanche events. Previous experiments have shown that the applied shear strain rate is sufficiently slow to separate individual avalanches and avoid avalanche overlap \cite{Denisov16}. This enables us to extract a wide range of predicted scaling exponents and scaling functions that identify the underlying slip statistics and dynamics \cite{Denisov16,Denisov17}. The second granular system, a rotary shear cell ( see Methods section), similarly shows strong intermittency: stress slowly increases, followed by sudden stress drops that correspond to acoustic emission (AE) events (Fig.\ref{figure_stress}~middle).  
We also analyze numerical simulations of a two-dimensional emulsion using Lattice Boltzmann Equation (LBE) modelling \cite{LBE1} (see Methods section). The model aims at simulating two repelling fluids. Coarsening is strongly suppressed by using a frustration mechanism, which stabilizes the interface. The system exhibits a yield stress rheology with a non-Newtonian relation between stress and shear strain rate above the yield stress \cite{LBE2}. For a small imposed shear strain rate, such that the stress is below yield stress, a clear stick-slip behaviour is observed (Fig. \ref{figure_stress}~bottom).

\subsection*{Avalanche size distributions}

\begin{figure}
\centering
\includegraphics[width=\linewidth]{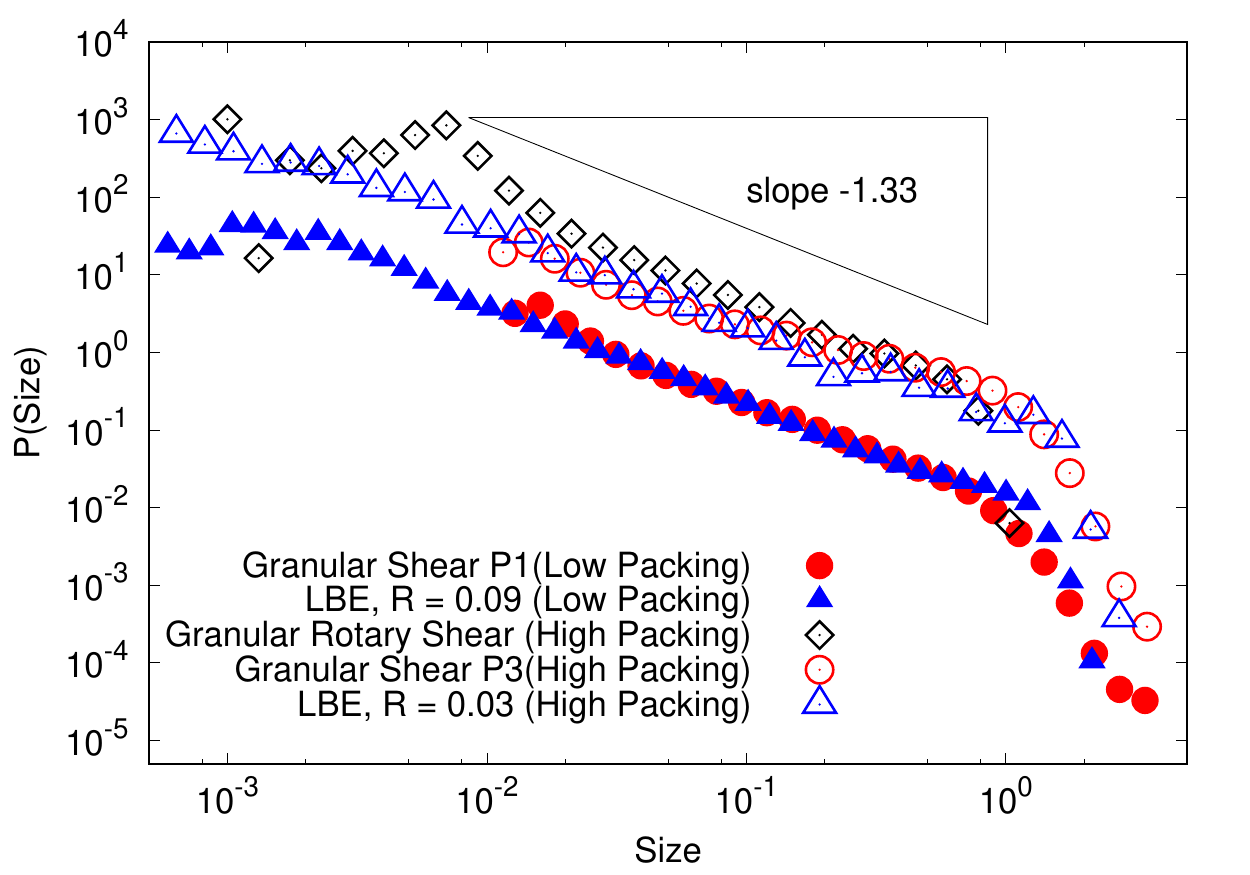}
\caption{Scaling properties of avalanche sizes measured for the granular systems (red circles and black diamonds) and the LBE simulations (blue triangles). Filled symbols indicate low packing fractions ($\phi=0.55$ for the granular material and $R=0.09$ for the LBE simulations), while open symbols indicate high packing fractions ($\phi=0.6$, respectively $0.64$, for the granular experiments and $R=0.03$ for the LBE simulations). The probability distributions corresponding to the high packing fractions have been shifted upwards for readability. }
\label{figure_size}
\end{figure}

The scaling properties of the avalanche sizes are similar for all our systems (figure \ref{figure_size}). Results from two different packing fractions are shown:  The distributions clearly collapse on top of each other, and all systems show a scaling with an exponent of $-1.33$ over 2-3 orders of magnitude in avalanche size, irrespective of their packing fraction. 

\subsection*{Avalanche interevent time distributions}

\begin{figure}
\centering
\includegraphics[width=\linewidth]{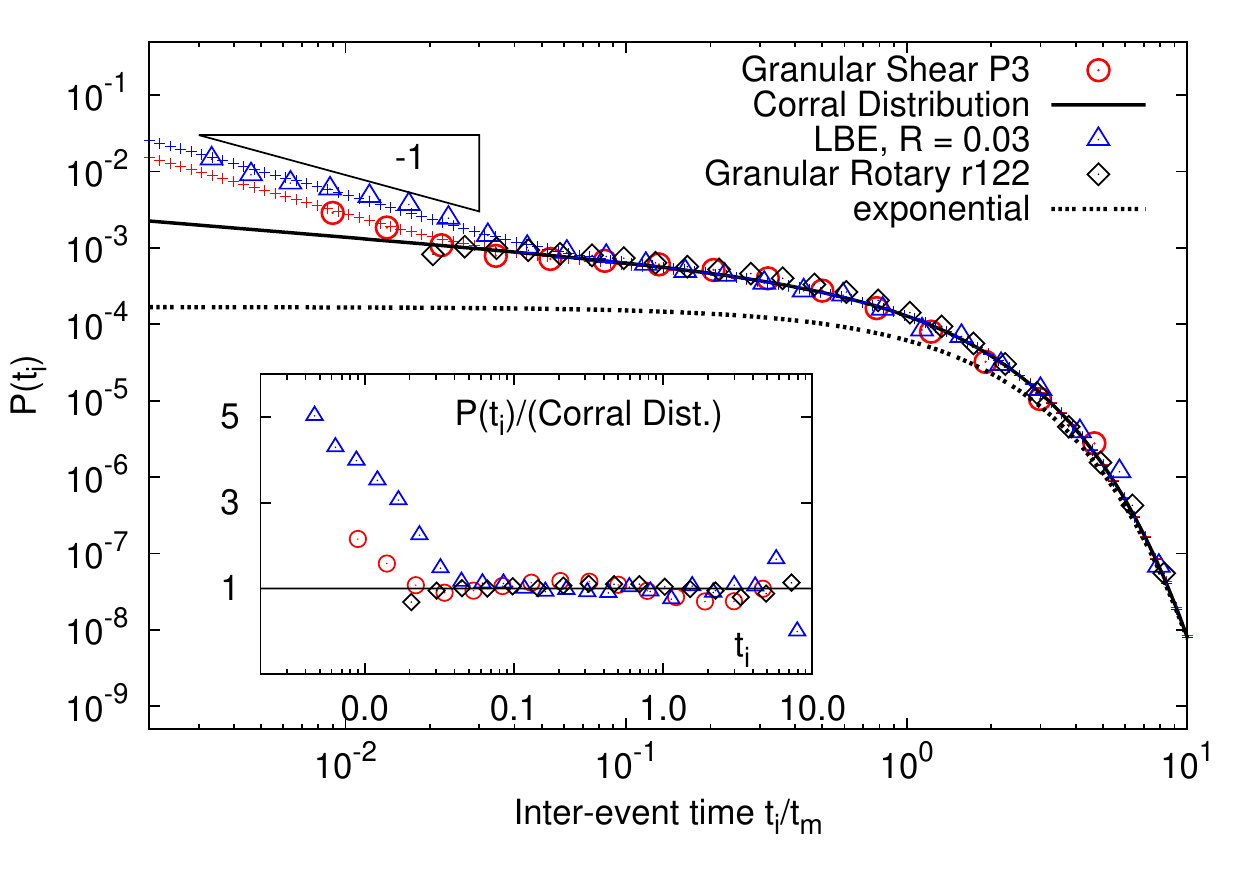}
\caption{Probability distribution $P(t_i)$ of interevent time $t_i$ between two consecutive avalanches in different physical systems at high packing fractions. The blue triangles correspond to numerical simulations of soft glasses while the red circles and black diamonds refer to laboratory results from three dimensional granular systems. The black line is the Corral best fit (see eq. (\ref{1})) obtained from the interevent time distribution for earthquakes. The dotted line is an exponential to mark the clear difference with the Gamma distribution. The red and blue dotted lines indicate the slope $-1$. The rotary granular experiment lacks  time resolution at short times.
Insert: Ratio $R(t_i)$ between the probability distributions $P(t_i)$ plotted in main figure with the one given by eq. (\ref{1}) with $ \gamma = 0.7$.}
\label{figure_high}
\end{figure}

The scaling properties of interevent times, however,  behave markedly  differently depending on packing fractions. 
At high packing fraction, all data collapse onto a Gamma distribution given by 
\begin{equation}
G(t_i)=\frac{C}{t_m}(\frac{t_i}{t_m})^{(\gamma-1)}\exp(-\frac{t_i}{t_m})
\label{1}
\end{equation} 
with $\gamma =0.7$, $t_m=\langle t_i \rangle$ the average interevent time and $C$ a normalization constant. This is the same Gamma distribution as that observed for earthquakes \cite{Corral2004}. The quality of the fitting is shown in the inset of Fig.~\ref{figure_high} where we plot the ratio $R(t_i) \equiv P(t_i)/G(t_i)$ where $P(t_i)$ are the different probability distributions shown in the figure and $G(t_i)$ is given by Eq. (\ref{1}).  
Furthermore, for short interevent times, most data 
display a clear
scaling behavior $P(t_i) \sim 1/t_i$ consistent with Omori's law. 
The short time regime can also be collapsed onto the same Gamma distributon after proper rescaling \cite{Corral2004}, but this is not essential for our main findings.

\begin{figure}
\centering
\includegraphics[width=\linewidth]{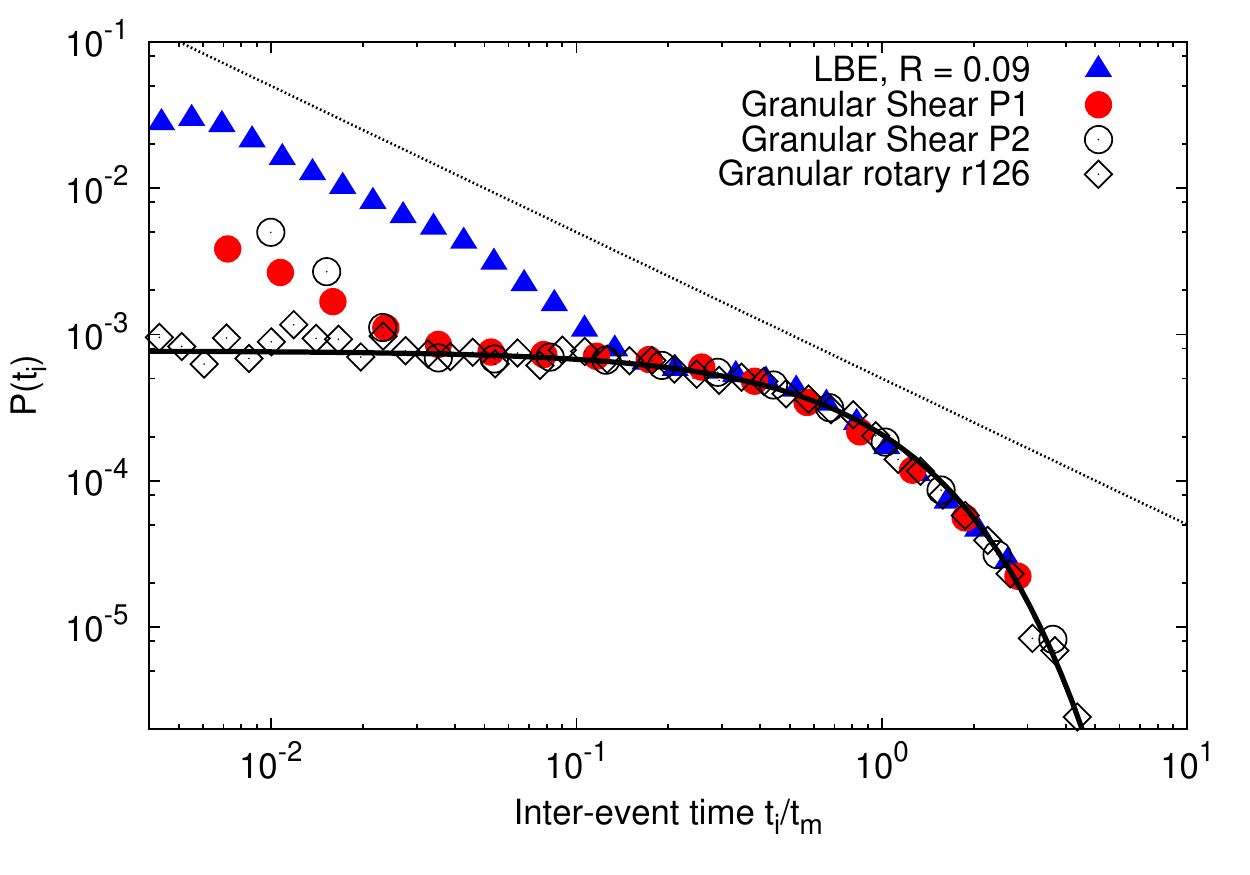}
\caption{Probability distribution $P(t_i)$ of the interevent time $t_i$ between two consecutive avalanches in different physical systems at low packing fractions. The blue triangles correspond to numerical simulations of soft glasses, while the circles refer to laboratory results from three dimensional granular systems and the diamonds to rotary experiments. The (black) dotted line indicates the slope $-1$ (Omori's behaviour is not seen for rotary experiment r126 on the times shown) and the (black) solid line an exponential curve. }
\label{figure_low}
\end{figure}

While earthquake data do not allow us to explore a wide range of packing fractions, the laboratory experiments and numerical simulations can easily explore lower packing fractions. For data corresponding to low packing fractions,  
$P(t_i)$ behaves markedly differently (Fig.~\ref{figure_low}). Two interesting features are seen in the probability distributions. For long $t_i/t_m$, the systems show an exponential behavior, corresponding to $\gamma=1$ in eq. (\ref{1}), clearly distinct from the distributions displayed in Fig.~\ref{figure_high}. The short time behavior $P(t_i) \sim 1/t_i$ is observed in all cases, except the rotary experiment r126, but no data collapse is obtained, i.e. the time scale separation between Omori's law and the exponential behavior is expressed differently depending on the physical system. 

Figs. (\ref{figure_high} and \ref{figure_low}) demonstrate the main point of our paper: $P(t_i)$ changes upon varying the system properties, i.e.the packing fraction, although no relevant change is observed in the avalanche size distribution (Fig. \ref{figure_size}). As the estimation of $t_i$ is done with the same methodology in all cases, the change of $P(t_i)$ with packing fraction cannot be considered an artifact due to the chosen thresholds \cite{Santucci2016}.  Rather, since the interevent times inform us on the relaxation time of the system, and their distribution on the relaxation process, a change of $P(t_i)$ indicates a fundamental change in the nature of the  underlying relaxation processes. In this context, it is noteworthy that at high packing fraction, all interevent time distributions are the same as those observed for earthquakes \cite{Corral2004,Corral2006,Saichev2006,Davidsen2}, suggesting a similar relaxation mechanism for earthquakes.

\section*{Discussion}

From the above, we can draw some general and nontrivial conclusions. 

\begin{figure}
\centering
\includegraphics[width=\linewidth]{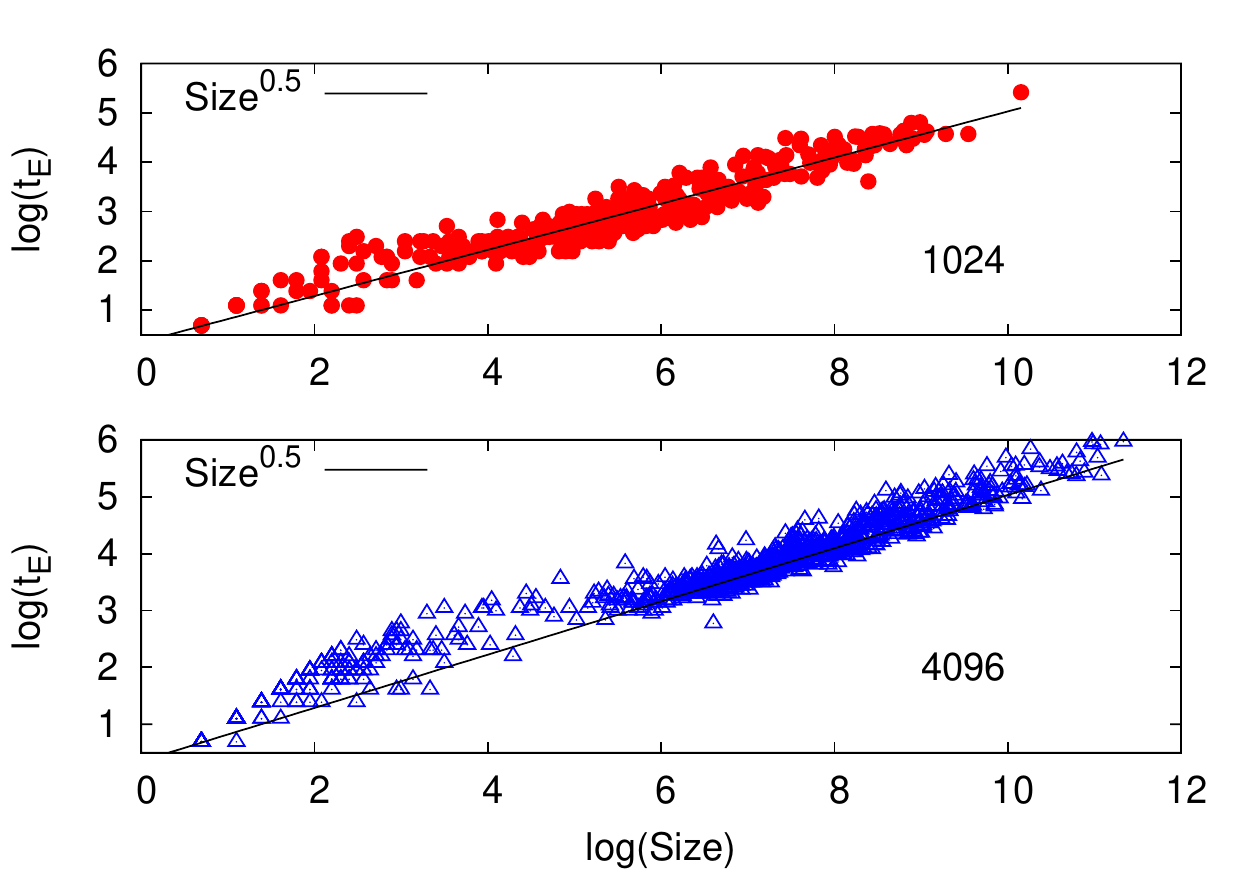}
\caption{Upper panel: avalanche time duration $t_E$ versus avalanche size $S$ observed in LBE simulation $L=1024, R=0.09$. Bottom panel: the same as above but for LBE simulation $L=4096, R=0.03$. In both simulations, the duration time $t_E$ of the avalanche satisfies the scaling relation $ t_E \sim S^z $ with $z \sim 1/2 $.}
\label{figure_duration}
\end{figure}

The statistical properties of avalanche size, $S$, and duration, $t_E$ (Figs. \ref{figure_size} and \ref{figure_duration}) \cite{Denisov17}, are independent of the system stiffness (i.e. the packing fraction). Physically this means that, once the system starts to release elastic energy, it does this independently of the system's packing fraction. This is one of the basic assumptions in many theoretical frameworks so far proposed to predict the (universal) scaling properties of avalanche size distribution.

The interevent time distributions, however, show a clear dependency on packing fractions (Figs. \ref{figure_high} and \ref{figure_low}). They also display a clear signature of two different time scales shaping their probability distribution. At short time scales, we observe events clustered in time where the interevent time distribution is $P(t_i)\sim 1/t_i$, regardless of packing fraction. This is consistent with Omori's law observed for earthquakes, and is most likely due to smaller size events triggered by some master event (after-shocks). Note, however, that our analysis is independent of any definition of main- and after-shock. At longer time scales, the interevent time distribution is given by Eq. (\ref{1}), where the exponent $\gamma$ depends clearly on packing fraction. At low packing fractions $\gamma$ is close to 1, and at high packing fractions it is about 0.7, close to the value found for earthquakes.
It is also important to remark that different experiments don't superimpose upon renormalization at low packing fractions, while the data collapse on the same functional form at high packing fractions. While we are not aware of any theoretical framework able to describe these features and/or explain how $P(t_i)$ changes with material stiffness, it has been argued \cite{Corral2006} that for Eq.~(\ref{1}) to emerge with $\gamma \ne 1$, correlations have to be present in the system. The question thus is whether our systems display any correlations. 

\begin{figure}
\centering
\includegraphics[width=\linewidth]{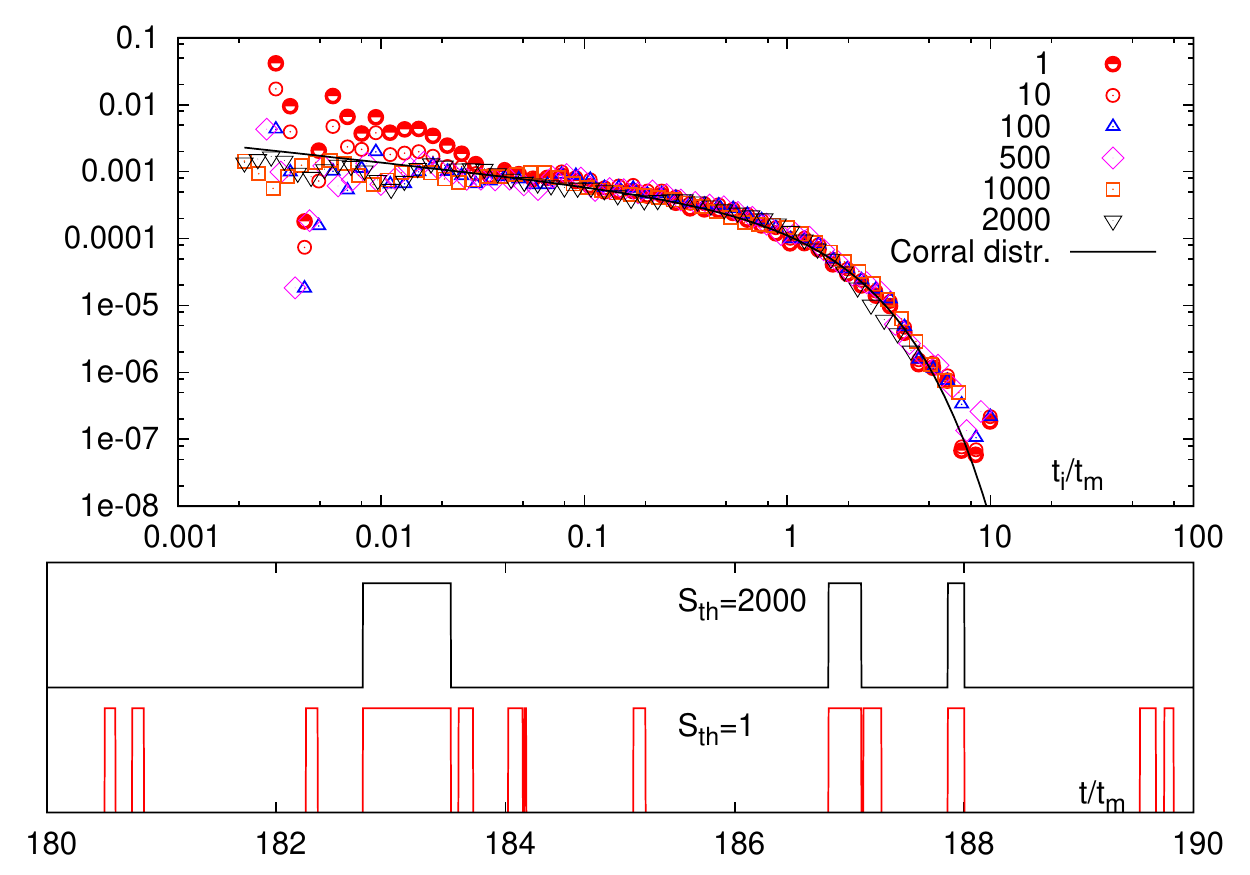}
\caption{Top panel: $P(t_i/t_m, S_{th})$ for $S_{th} \in [1,2000]$. In the range $S_{th} \in [1,2000]$ the probability distributions $P(t_i/t_m, S_{th})$ collapse on the same master curve for $t/t_m \ge 0.03$, which is close to the Gamma distribution with $\gamma=0.7$. Bottom panels: event snapshots for the two extreme cases $S_{th}=1$ and $S_{th}=2000$. The former displays indeed many independent events from the latter.}
\label{figure_threshold}
\end{figure}

Fig. (\ref{figure_size}) demonstrates that our systems display strong correlations in space \cite{Denisov16,Barrat2018}. To investigate whether or not any correlations exist in time (''memory effect'' in the avalanche dynamics), we follow Corral \cite{Corral2005} and consider the probability density of the interevent times conditioned on the avalanche size. We focus of the numerical simulation at high packing fraction ($R=0.03$) and define $P(t_i/t_m, S_{th})$ as the interevent time probability density for a subset of events with size $S > S_{th}$. Upon increasing $S_{th}$, if $P(t_i/t_m, S_{th})$ is not exponentially distributed and it does not change its shape, then time correlations (or memory effects) exist between the interevent times for different threshold values $S_{th}$. For a process with no correlations, one can show that the only scale invariant distribution is the exponential distribution \cite{Molchan}. On the other hand, correlations introduce new invariant functions in the process. The robust shape of the distributions for different $S_{th}$ in Fig. (\ref{figure_threshold}) then clearly demonstrates that the interevent time is indeed correlated in a non-trivial way with the size of the previous event.

Motivated by this observation, we further looked for spatio-temporal correlations by analyzing sub-regions of the system. We consider two disjoint regions, say $A$ and $B$, and their set union, $A \cup B$ including both regions, over the full simulation time interval. If the regions are independent, the only possible interevent time distribution is exponential \cite{Molchan}. If there are space-time correlations between the different regions, then we expect that the probability distributions $P(t_i/t_m(A))$, $P(t_i/t_m(B))$ and $P(t_i/t_m(A \cup B))$ are all the same and none of them is exponential \cite{Corral2005, Corral2006}. If at least  in one of the regions $A$, $B$ and $A \cup B$  the interevent time is exponentially distributed, then we can argue that the Gamma distribution observed for the whole system is just by accident.  This constitutes a severe test to uncover (although indirectly) space-time correlations in the system.

\begin{figure}
\centering
\includegraphics[width=\linewidth]{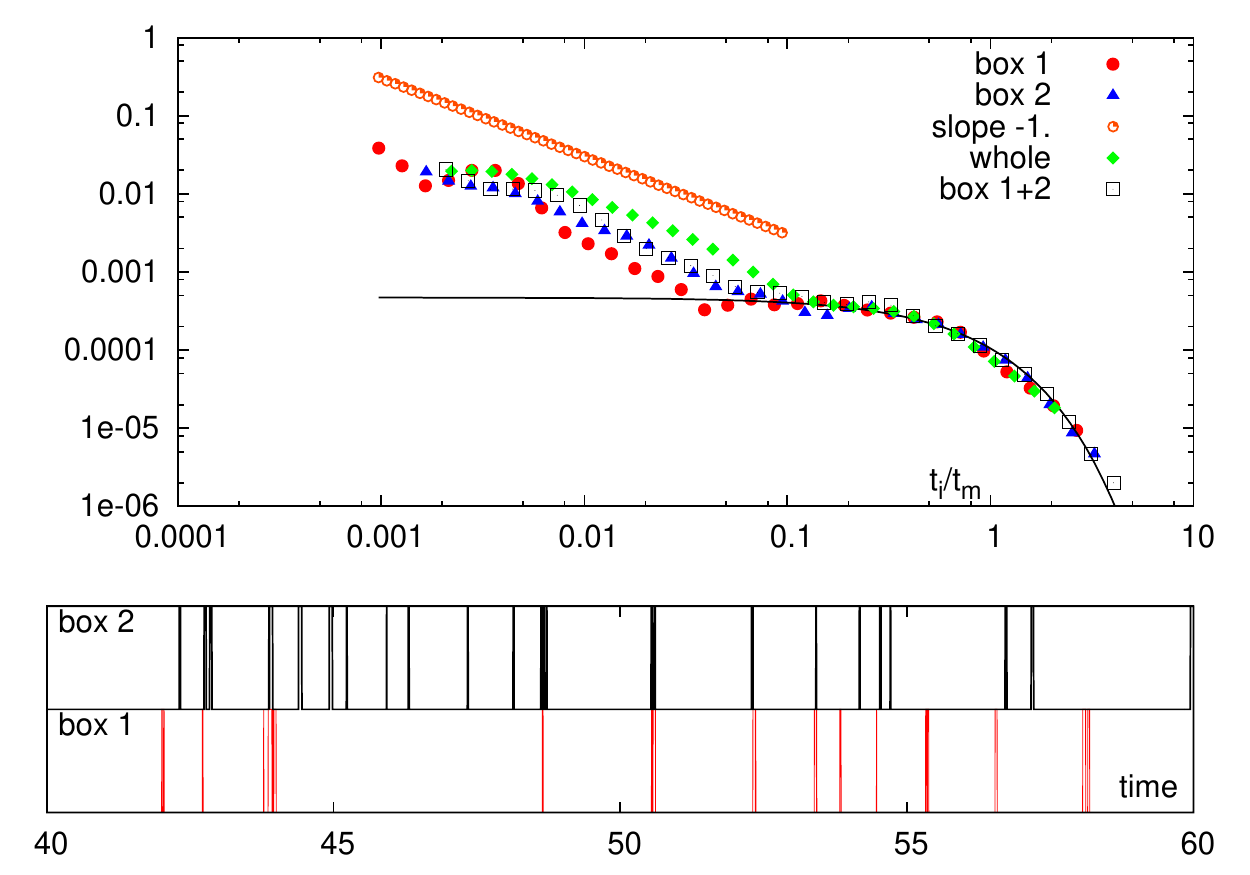}
\caption{Top panel: $P(t_i/t_m)$ for the different boxes, their sum and the whole system for the case $R=0.09$.  Bottom panels: event snapshots showing the independence of the 2 time series. }
\label{figure_box_low}
\end{figure}

In the original picture of avalanche dynamics in amorphous systems, the usual assumption is that while the avalanches themselves are highly correlated events, they occur at random uncorrelated times i.e. with an exponential distribution for interevent times $P(t_i/t_m)$. For such uncorrelated random events, we expect therefore that the interevent time distributions $t_i/t_m(A)$ and $t_i/t_m(B)$ are both also exponentially distributed. 
This is what happens for our LBE simulation at relatively low packing fraction ($R=0.09$) as shown in the upper panel of figure (\ref{figure_box_low}). All curves follow the exponential distribution. In the lower panel of the same figure we show a snapshot of the time series illustrating the avalanche events in the two regions (referred to as $box \, 1$ and $box \, 2$). In the upper panel we also show the interevent time distribution for the whole system, which, as we know, is also exponential (Fig. \ref{figure_low}). Note that the curves do not collapse, as they apparently sense the short time scale differently.

\begin{figure}
\centering
\includegraphics[width=\linewidth]{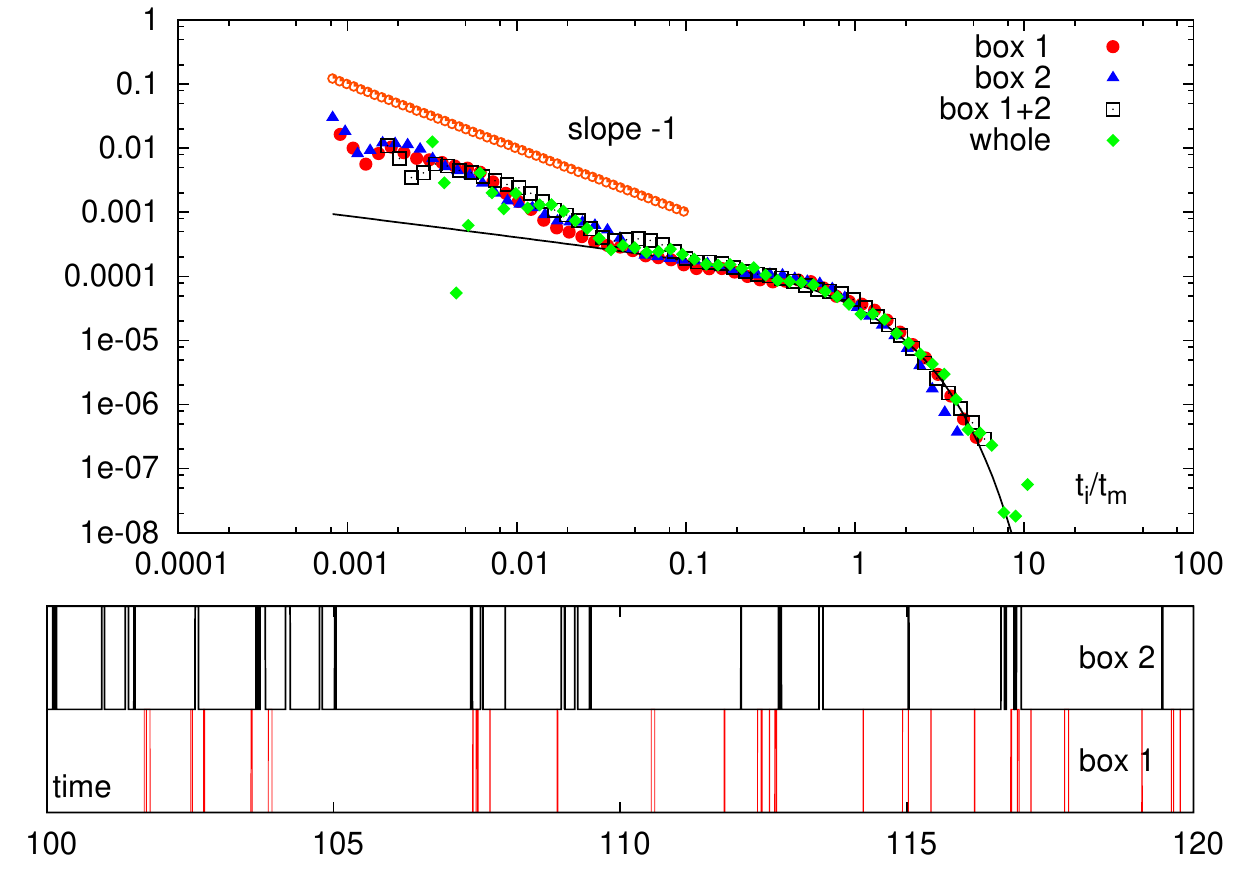}
\caption{Top panel: $P(t_i/t_m)$ for the different boxes, their sum and the whole system for the case $R=0.03$. Bottom panels: event snapshots. Bottom panels: event snapshots showing the independence of the 2 time series.}
\label{figure_box_high}
\end{figure}

The same analysis for the case of a large packing fraction $R=0.03$) gives a different picture (Fig. \ref{figure_box_high}). In all cases, we indeed observe the same probability distribution for $P(t_i/t_m)$, which is not exponential. As we have argued above, this can only be true if there are non-trivial correlations and/or memory effects between the two different regions. Note that in this case the curves collapse also for the short Omori time scales.

While it is hard to identify the exact nature of such correlations, a clear picture merges from our observations. Overall our results demonstrate that the interevent time distribution $P(t_i)$ is able to disentangle the statistical properties of systems at different packing fractions, whereas this is not possible by looking only at the avalanche size distribution and duration. As the interevent times are directly related to the relaxation time of the system, their distribution should contain information about the relaxation mechanism. We thus conclude that the different interevent time distributions we observe, and their interpretation in terms of absence and presence of memory effects, indicate fundamentally different relaxation mechanisms of the systems at these different volume fractions or confining pressures.


\section*{Methods}

\subsection*{Laboratory Experiments}

\begin{figure}
\centering
\includegraphics[width=\linewidth]{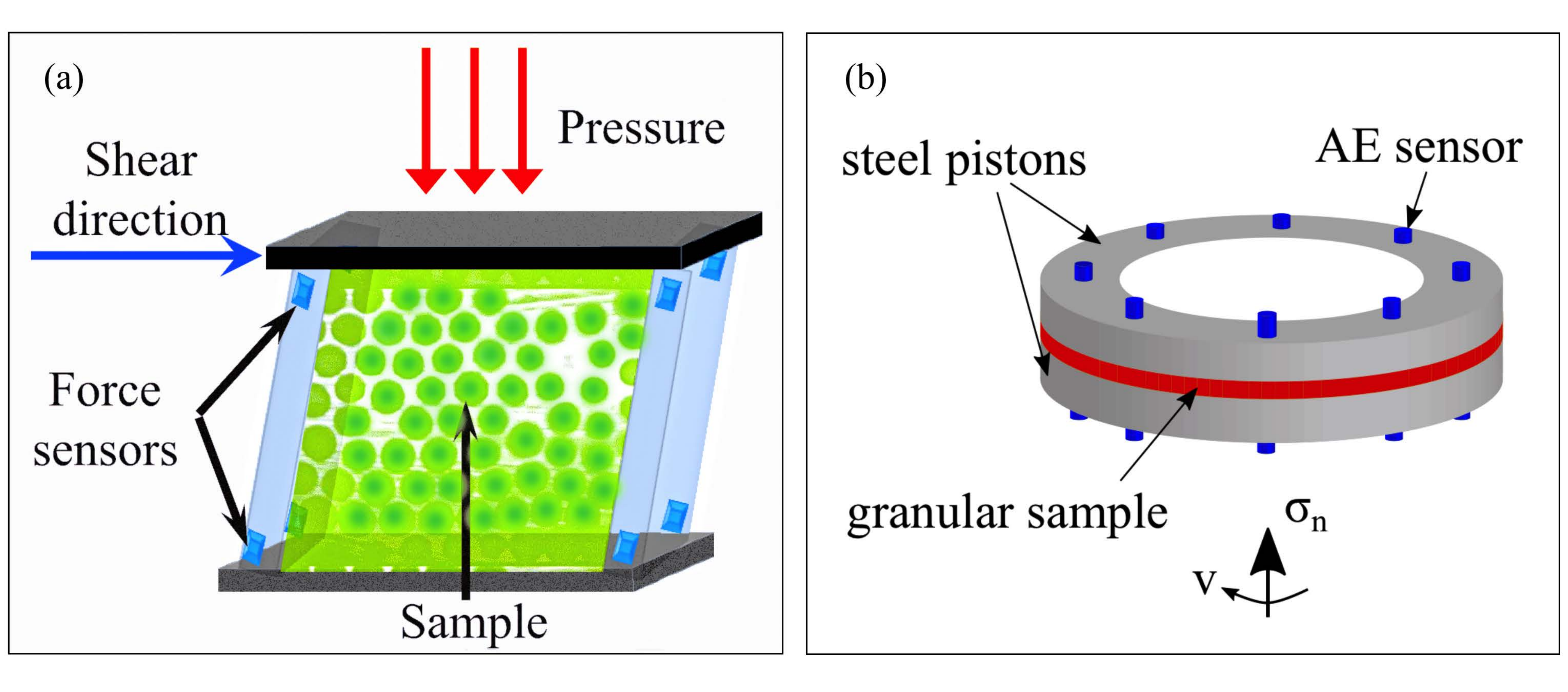}
\caption{ Figures (a) and (b) schematically show the granular shear cells with force or acoustic emission (AE) sensors in the walls. Loads imposed perpendicular to shear exert a constant confining pressure.}
\label{figure_granular}
\end{figure}

We use two different granular systems, where we apply well-controlled normal forces and slow shear strain rates (Fig. \ref{figure_granular}). 

The first system, a shear cell, consists of $3\cdot10^5$ polymethyl methacrylate spheres with a diameter of $d=1.5$ mm and a polydispersity of $\sim5\%$, and is deformed at a constant shear strain rate of $\dot{\gamma}=9.1 \cdot 10^{-4}$ $s^{-1}$ and constant normal stress, using a shear cell with a confining top plate (Fig. \ref{figure_granular}a) \cite{Denisov17}. By placing weights onto the confining plate, we vary the normal {stress} between 4, 6.8 and 9.6 kPa, experiments P1, P2 and P3, respectively, resulting in a packing fraction $\phi$ between $55\%$ and $60\%$. We monitor shear-induced force fluctuations using force sensors included in the shearing walls. 

The second granular system, a rotary shear cell, consists of glass bead layers that are sheared using a servo-controlled rotary shear apparatus (Fig. \ref{figure_granular}b). The average particle size is 0.5 mm and the standard deviation 0.1 mm. For each experiment, an approximately 4.5 mm thick layer of glass beads is deposited in an annular-shaped shear cell that consists of four independent steel rings: two of them act as pistons that provide vertical confinement, whereas the other two provide lateral support. The cell is then placed in the rotary shear apparatus, built inside an Instron 8862 testing machine. The servo-controlled Instron actuator is used to prescribe a constant normal stress condition (8 MPa for experiment r122 and 2 MPa for experiment r126), and a Parker MH-205 motor to maintain a constant angular velocity of $0.02 ~^{\circ}/s$ by rotating the bottom piston. An axially mounted load cell ($\pm 100$ kN range, 0.008 kN resolution) measures the normal stress, and a pair of laterally mounted load-cells (each 20 kN maximum load, 0.008 kN resolution) the traction. The experiments are started at a random close packing although there is an increase over time due to the breaking of particles. For the purpose of this paper, it is important to note that a confining pressure of 2 MPa is a proxy for a low packing fraction and 8 MPa for a high packing fraction experiment. AE activity is monitored via two arrays of 8 piezoelectric transducers each, mounted inside the two steel piston rings at $45^\circ$ intervals.  

\subsection*{Numerical Simulations}

\begin{figure}
\centering
\includegraphics[width=\linewidth]{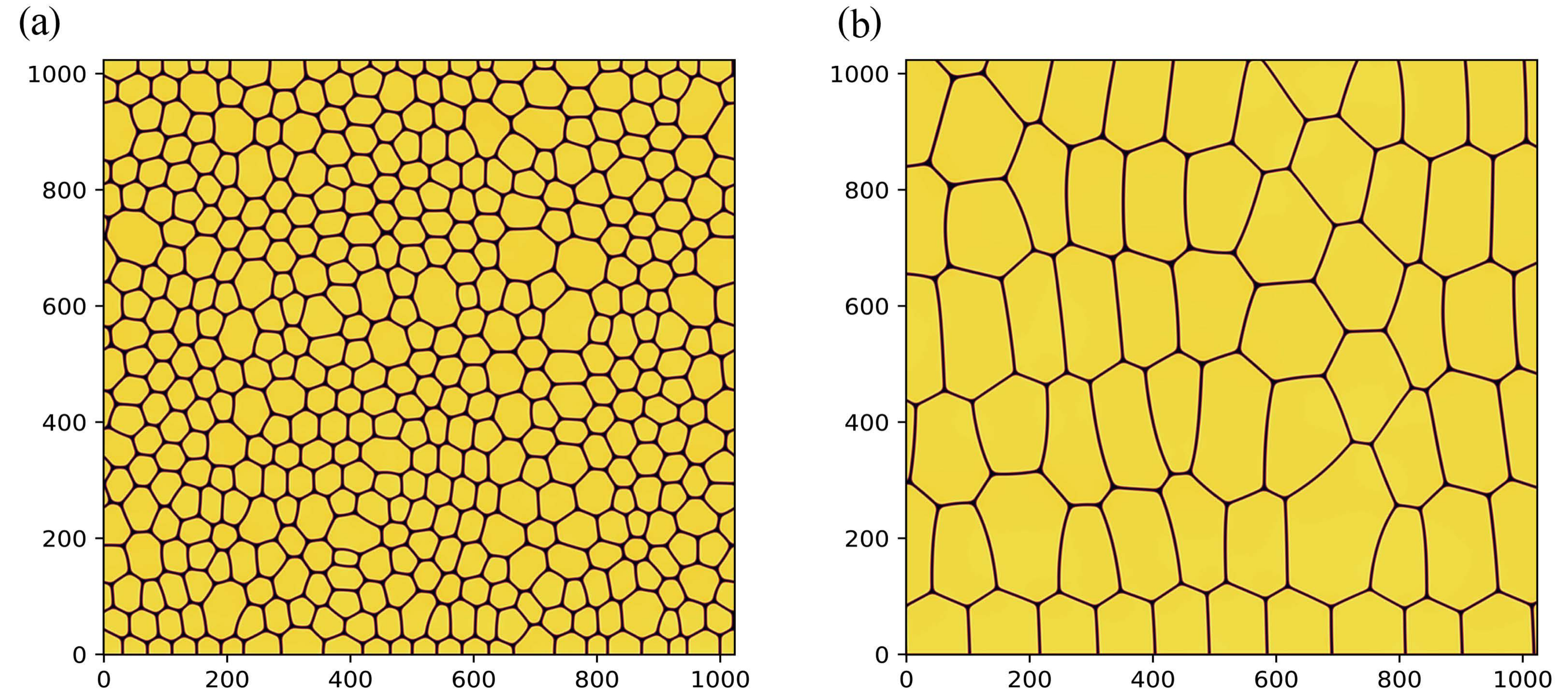}
\caption{The left panel corresponds to the case of LBE size $L=1024$ and $R=0.09$. The right panel shows a $1024^2$ portion of the simulation performed with $L=4096$ and $R=0.03$.}
\label{figure_emulsion}
\end{figure}

\begin{figure}
\centering
\includegraphics[width=\linewidth]{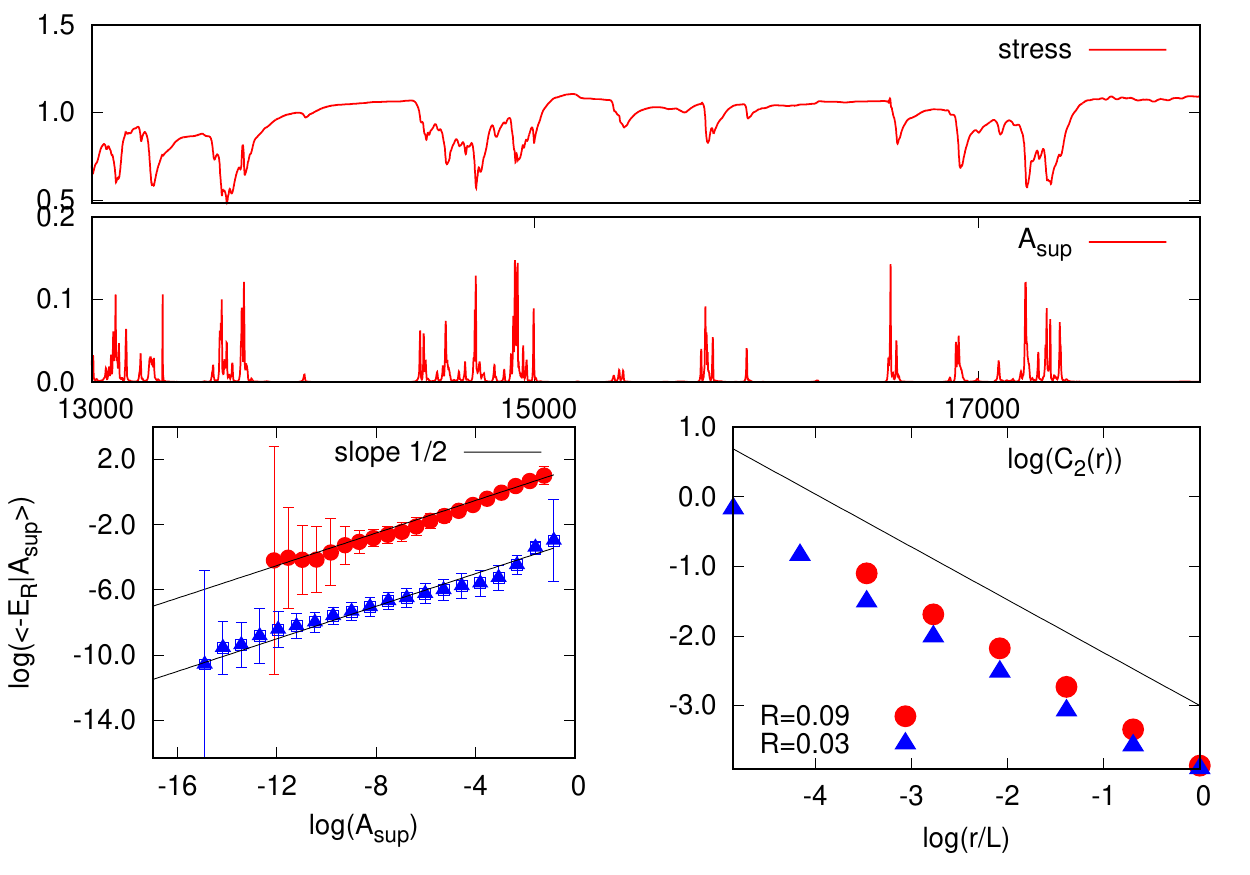}
\caption{Upper two panels. Top: snapshot of the $xy$ component of the stress $\sigma$ as a function of time. Bottom: behavior of the quantity $A_{sup}(t)$, defined in the text, for the same time interval. Note that stress drops correspond to relatively large values of $A_{sup}$ due to irreversible lattice rearrangements. Lower two panels. Left: the average energy release  $E_r$ conditioned on $A_{sup}$, see equation (\ref{4}), as a function of $A_{sup}$ for two LBE simulations. Right: the scaling behavior of $C_2(r)$ as a function of $r$ for the two numerical simulations with $R=0.09$ (red bullets) and $R=0.03$ (blue triangles) corresponding to LBE grids $1024$ and $4096$ respectively. }
\label{figure_simul}
\end{figure}

The simulations use a model based on Lattice Boltzmann equations (LBE) for complex fluids\cite{LBE1}, which are discussed in detail in \cite{LBE2,LBE3,LBE4}. The model simulates two repelling fluids, say $A$ and $B$, with the same density. Coarsening is strongly suppressed by using a frustration mechanism, which stabilizes the interface. The initial configuration is chosen such that $N$ droplets of the fluid $A$ are randomly created in space with small polydispersity. The interface is filled by fluid $B$. The system exhibits a yield stress rheology with a non-Newtonian relation between stress and shear strain rate above the yield stress \cite{LBE2}. The ratio $R$ between the interface area $A_{int}$ and the bubble area $A_b$ can be estimated as

\begin{equation}
\label{2}
R  \sim \frac{ 2 \delta \sqrt{N}}{L}
\end{equation}

where $\delta$ is the interface thickness, $N$ is the overall number of bubbles and $L$ is the size of the system in LBE units.  Note that, qualitatively, the quantity $1-R$ can be considered as the packing  fraction of the system. Since the interfaces are not sharp in our system, the correct packing fraction should be $1- C R$, with $C$ a constant of order $1$. Since $\delta$ must be finite for the interface to be stable, the only way to decrease $R$ (i.e. to increase the packing fraction) is to decrease the ratio $\sqrt{N}/L$. By decreasing $R$, we also increase the value of the yield stress, i.e. we increase the rigidity of our system.  Examples of initial configurations with $R = 0.09$ and $R=0.03$ are shown in figure (\ref{figure_emulsion}a) and (\ref{figure_emulsion}b). The initial conditions $R=0.03$ together with $L=4096$ give a similar number of bubbles to that obtained for $R=0.09$ and $L=1024$.  We will use results from both cases. The rheological properties of such systems are discussed in detail in~\cite{LBE2}. We perform simulations with a small externally imposed shear strain rate, whose value is chosen such that the stress is below the yield stress transition. An example of the resulting stress $\sigma$ as a function of time is shown in figure (\ref{figure_simul}) upper panel: a clear stick-slip behavior is observed. To perform a detailed statistical analysis of the dynamics, we used the method recently developed in \cite{GJI}. We consider $n^2$ small squares of size $L/n$ and, for each of the squares, we compute the quantity $A_i \equiv \langle (\rho_A(x,y,t+\tau)-\rho_A(x,y,t))^2 \rangle $, where $\langle ... \rangle$ is the space average over the square $i$ and $i=1,2,\dots,n^2$. We chose $n = 32$ and $\tau=1000$ LBE time steps. We checked that different choices do not change the results discussed in the rest of this section. The quantity $A_i$ is a measure of the relative number of points in square $i$ which move in time interval $\tau$, i.e. $A_i$ is the square of the displacement occurring in the square $i$. Plastic events are localized in space and correspond to the largest value of $A_i$ observed in the system at time $t$. Therefore the relevant quantity to consider is \cite{GJI}:

\begin{equation}
A_{sup}(t) = sup_i [A_i(t)]
\label{3}
\end{equation}

A large value of $A_{sup}(t)$ corresponds to a large stress drop in the system. In figure (\ref{figure_simul}) we show in the upper panel $\sigma(t)$ and in the middle panel $A_{sup}(t)$ for the same time windows. To make a more quantitative analysis, we computed  the quantity

\begin{equation}
\label{4}
E_r = [  -\sigma \frac{d\sigma}{dt} | A_{sup} ]
\end{equation}

where $E_r$ represents the time averaged value of the energy release $-\sigma d \sigma /dt$ conditioned on a particular value of $A_{sup}$. In the lower left panel of figure (\ref{figure_simul}) we show $E_r$ as a function of $A_{sup}$ for the two simulations at $R=0.09$ (red circles) and $R=0.03$ (blue triangles). A clear scaling law  with a slope of $1/2$ is observed. Using this result,  we can state that  the following relation holds scaling wise

\begin{equation}
\label{5}
E_r \sim A_{sup}^{1/2}
\end{equation}

Next we investigated the spatial correlation in the system and computed the variables

\begin{equation}
\label{psi}
\psi_r  \equiv \frac{1}{r^2} \int_{B(r)} dx dy \left[ \frac{A_i}{\langle A \rangle } \right]^2
\end{equation}

where $B(r)$ is a box of side $r$ and $ \langle ... \rangle$ denotes the spacial average. Using $\psi_r$, we can construct the multi-fractal quantities $C_q(r) =  E[ { \langle \psi_r^q \rangle } ]$, where $E[...]$ is the time average. For a multi-fractal system we should observe $C_q(r) \sim r^{(q-1)(D_q-d)}$, where the exponents $D_q$ are the generalized fractal dimensions and $d$ is the space dimension ($d=2$ in our case). Using the above definition of $C_q$, the space correlation of $A_i$ is associated with $D_2$ known as the correlation dimension, namely $E [ \langle A_i A_j \rangle ] \sim r^{D_2-2} $ where $A_i$ ans $A_j$ are separated by the distance $r$. The red circles in the lower right panel of figure (\ref{figure_simul}) corresponds to the quantity $C_2(r)$ computed from $A_i$. A clear scaling is observed with exponent $ \sim -0.7$ corresponding to $D_2 \sim 1.3$. Therefore we can conclude that the system displays strong correlations in space. The same feature is observed using the overlap-overlap correlation function as discussed in \cite{LBE3}.

We then analyzed the probability distribution of $A_{sup}$. In Fig. (\ref{figure_size}), we show $P(A_{sup})$ as a function of $A_{sup}$ for $R=0.09$ and $R=0.03$. As already discussed  in \cite{GJI}, a clear scaling is  observed for  $A_{sup}> A_{th}$ over 2 decades, where $A_{th}$ is some threshold value (the same for both cases). The scaling exponent $P(A_{sup}) \sim A_{sup}^{-\gamma}$ is given by $\gamma \sim 1.33$ . The quality of the scaling is best appreciated in the inset where we compensate for the power law, i.e we plot $P(A_{sup}) / A_{sup}^{-1.33}$. The clear definition of $A_{th}$ appearing in the figure can be used to compute the scaling properties of the avalanche dynamics. We defined the time duration  $t_E$ of the avalanche as the time interval when $A_{sup}>A_{th}$. Within each avalanche of duration $t_E$, we computed the size $S$ of the avalanche as the number of the $n^2$ region where $A_i>A_{th}$. The numerical simulations were performed for $40 \times 10^6$ LBE steps in the case of $R=0.09$ and for $280 \times 10^6$ LBE times steps for the case of $R=0.03$.  A clear dynamical scaling $ t_E \sim S^{z}$ is observed with $z \sim 1/2$, close to mean field predictions \cite{Fisher}. From this dynamical scaling we therefore obtain

\begin{equation}
\label{6}
S \sim t_E^2 \sim \int_{t_E} dt [-\sigma \frac{d\sigma}{dt}] \sim t_E E_r  \sim t_E A_{sup}^{1/2} \sim A_{sup}
\end{equation}

It follows that the scaling exponent $\gamma$ previously defined is also the scaling exponent of the probability distribution of $S$, i,e, $P(S) \sim S^{-1.33}$ similar to what has been found in many numerical simulations, see the numerical values reported in \cite{Wyart}.

\bibliographystyle{plain}
\bibliography{interevent}

\begin{thebibliography}{10}

\bibitem{LBE2}
R.~Benzi, M.~Bernaschi, M.~Sbragaglia, and S.~Succi.
\newblock Herschel-bukley rheology from lattice kinetic theory of soft glassy
  materials.
\newblock {\em Europhys. Lett.}, 91(1):14003, 2010.

\bibitem{GJI}
R.~Benzi, P.~Kumar, F.~Toschi, and J.~Trampert.
\newblock Earthquake statistics and plastic events in soft-glassy materials.
\newblock {\em Geophys. J. Int.}, 207:1667--1674, 2016.

\bibitem{LBE3}
R.~Benzi, M.~Sbragaglia, M.~Perlekar, P.~Bernaschi, S.~Succi, and F.~Toschi.
\newblock Direct evidence of plastic events and dynamic heterogeneities in
  soft-glasses.
\newblock {\em Soft Matter}, 10:4615--4624, 2014.

\bibitem{LBE4}
R.~Benzi, M.~Sbragaglia, A.~Scagliarini, P.~Perlekar, M.~Bernaschi, S.~Succi,
  and F.~Toschi.
\newblock Internal dynamics and activated processes in soft-glassy materials.
\newblock {\em Soft Matter}, 11:1271--1280, 2015.

\bibitem{LBE1}
R.~Benzi, M.~Sbragaglia, S.~Succi, M.~Bernaschi, and S.~Chibbaro.
\newblock Mesoscopic lattice boltzmann modeling of soft-glassy systems: Theory
  and simulations.
\newblock {\em J. Chem. Phys.}, 131:104903, 2009.

\bibitem{Corral2004}
\'A. Corral.
\newblock Long-term clustering, scaling and universality in the temporal
  occurrence of earthquakes.
\newblock {\em Phys. Rev. Lett.}, 92:108501, 2004.

\bibitem{Corral2005}
\'A. Corral.
\newblock Time-decreasing hazard and increasing time until the next earthquake.
\newblock {\em Phys. Rev. E}, 71:017101, (2005.

\bibitem{Corral2006}
\'A. Corral.
\newblock Universal earthquake-occurrence jumps, correlations with time and
  anomalous diffusion.
\newblock {\em Phys. Rev. Lett.}, 97:178501, (2006.

\bibitem{BenZion}
K.~A. Dahmen, Y.~Ben-Zion, and J.~T. Uhl.
\newblock A micro-mechanical model for deformation in solids with universal
  predictions for stress-strain curves and slip-avalanches.
\newblock {\em Phys. Rev. Lett.}, 102:175501, 2009.

\bibitem{Davidsen2}
J\"orn Davidsen and Grzegorz Kwiatek.
\newblock Earthquake interevent time distribution for induced micro-, nano-,
  and picoseismicity.
\newblock {\em Phys. Rev. Lett.}, 110:068501, Feb 2013.

\bibitem{Davidsen}
J\"orn Davidsen, Sergei Stanchits, and Georg Dresen.
\newblock Scaling and universality in rock fracture.
\newblock {\em Phys. Rev. Lett.}, 98:125502, Mar 2007.

\bibitem{Denisov16}
D.~V. Denisov, K.~A. Lorincz, J.~T. Uhl, K.~A. Dahmen, and P.~Schall.
\newblock Universality of slip avalanches in flowing granular matter.
\newblock {\em Nat. Commun.}, 7:10641, 2016.

\bibitem{Denisov17}
D.~V. Denisov, K.~A. Lorincz, W.~J. Wright, T.~C. Hufnagel, A.~Nawano, X.~Gu, ,
  J.~T. Uhl, K.~A. Dahmen, and P.~Schall.
\newblock Universal slip dynamics in metallic glasses and granular matter --
  linking frictional weakening with inertial effects.
\newblock {\em Sci. Rep.}, 7:43376, 2017.

\bibitem{Fisher}
D.~S. Fisher.
\newblock Collective transport in random media: from superconductors to
  earthquakes.
\newblock {\em Physics Reports}, 301(1):113 -- 150, 1998.

\bibitem{gb:1954}
B.~Gutenberg and C.F. Richter.
\newblock {\em Seismicity of the earth and associated phenomena}.
\newblock Princeton University Press, Princeton, 1954.

\bibitem{Santucci2016}
Sanja Jani\ifmmode \acute{c}\else \'{c}\fi{}evi\ifmmode~\acute{c}\else
  \'{c}\fi{}, Lasse Laurson, Knut~J\o{}rgen M\aa{}l\o{}y, St\'ephane Santucci,
  and Mikko~J. Alava.
\newblock Interevent correlations from avalanches hiding below the detection
  threshold.
\newblock {\em Phys. Rev. Lett.}, 117:230601, Dec 2016.

\bibitem{Goldenfeld}
M.~LeBlanc, L.~Angheluta, K.~Dahmen, and N.~Goldenfeld.
\newblock Universal fluctuations and extreme statistics of avalanches near the
  depinning transition.
\newblock {\em Phys. Rev. E}, 87:022126, 2013.

\bibitem{Wyart}
J.~Lin, E.~Lerner, A.~Rosso, and M.~Wyart.
\newblock Scaling description of the yielding transition in soft amorphous
  solids at zero temperature.
\newblock {\em Proc. Natl. Acad. Sci.}, 111:14382--14387, 2014.

\bibitem{maass_etal:2015}
R.~Maa{\ss}, M.~Wraith, J.~T. Uhl, J.~R. Greer, and K.~A. Dahmen.
\newblock {Slip statistics of dislocation avalanches under different loading
  modes}.
\newblock {\em Phys. Rev. E}, 91(042403):doi:10.1103/PhysRevE.91.042403, 2015.

\bibitem{Zapperi}
M.~C. Miguel, A.~Vespignani, S.~Zapperi, J.~Weiss, and J.~R. Grasso.
\newblock Intermittent dislocation flow in viscoplastic deformation.
\newblock {\em Nature}, 410:667--671, 2001.

\bibitem{Molchan}
G.~Molchan.
\newblock Interevent time distribution of seismicity: a theoretical approach.
\newblock {\em Pure Appl. Geophys.}, 162:1135--1150, (2005.

\bibitem{Niccolini}
G.~Niccolini, A.~Carpinteri, G.~Lacidogna, and A.~Manuello.
\newblock Acoustic emission monitoring of the syracuse athena temple: Scale
  invariance in the timing of ruptures.
\newblock {\em Phys. Rev. Lett.}, 106:108503, Mar 2011.

\bibitem{Barrat2018}
Alexandre Nicolas, Ezequiel~E. Ferrero, Kirsten Martens, and Jean-Louis Barrat.
\newblock Deformation and flow of amorphous solids: Insights from elastoplastic
  models.
\newblock {\em Rev. Mod. Phys.}, 90:045006, Dec 2018.

\bibitem{Ribeiro}
H.~V. Ribeiro, L.~S. Costa, L.~G.~A. Alves, P.~A. Santoro, S.~Picoli, E.~K.
  Lenzi, and R.~S. Mendes.
\newblock Analogies between the cracking noise of ethanol-dampened charcoal and
  earthquakes.
\newblock {\em Phys. Rev. Lett.}, 115:025503, Jul 2015.

\bibitem{Saichev2006}
A.~Saichev and D.~Sornette.
\newblock ``universal'' distribution of interearthquake times explained.
\newblock {\em Phys. Rev. Lett.}, 97:078501, Aug 2006.

\bibitem{Sethna}
J.~P. Sethna, K.~A. Dahmen, and C.~R. Myers.
\newblock Crackling noise.
\newblock {\em Nature}, 410:242--250, 2001.

\bibitem{Touati}
Sarah Touati, Mark Naylor, and Ian~G. Main.
\newblock Origin and nonuniversality of the earthquake interevent time
  distribution.
\newblock {\em Phys. Rev. Lett.}, 102:168501, Apr 2009.

\bibitem{uhl_etal:2015}
J.~T. Uhl, S.~Pathak, D.~Schorlemmer, X.~Liu, R.~Swindeman, B.~A.~W. Brinkman,
  M.~LeBlanc, G.~Tsekenis, N.~Friedman, R.~Behringer, D.~Densiov, P.~Schall,
  X.~Gu, W.~J. Wright, T.~Hufnagel, A~Jennings, J.~R. Greer, P.~K. Liaw,
  T.~Becker, G.~Dresen, and K.~A. Dahmen.
\newblock {Universal quake statistics: From compressed nanocrystals to
  earthquakes}.
\newblock {\em Scientific reports}, 5(16493):doi:10.1038/srep16493, 2015.

\bibitem{utsu_etal:1995}
T.~Utsu, Y~Ogata, and R.S. Matsu'ure.
\newblock The centenary of the omori formula for a decay law of aftershock
  activity.
\newblock {\em J. Phys. Earth}, 43:1--33, 1995.

\end{thebibliography}

\section*{Acknowledgements}

This research was partly funded by the Shell-NWO/FOM programme `Computational sciences for energy research' under project number 14CSER022, partially funded by ERC Starting Grant 335915 (SEISMIC) and partially funded by NWO VIDI grant 854.12.011.  Numerical simulations for this work were carried out on the Dutch national e-infrastructure with the support of SURF Cooperative.

\section*{Author contributions statement}
R.B., A.N., P.S., F. T. and J.T. conceived the study,  P.K., E.K. and D.D. conducted the experiments, R.B. analysed the results. All authors reviewed the manuscript. 

\end{document}